\documentclass[letterpaper, 10 pt, journal]{IEEEtran}
\IEEEoverridecommandlockouts                          
\usepackage{graphics} 
\usepackage{epsfig}
\usepackage{times}
\usepackage{amsmath}
\usepackage{amssymb}
\usepackage{threeparttable}
\usepackage{booktabs}
\usepackage{soul}
\usepackage{subfigure}
\usepackage{xcolor}
\usepackage[hidelinks]{hyperref}
\usepackage{algorithm}
\usepackage{algorithmic,comment}
\usepackage{tcolorbox}
\usepackage{cite}
\usepackage[edges]{forest}
\usepackage{tikz}

\title{\LARGE \bf
Redefining End-of-Life: Intelligent Automation for Electronics Remanufacturing Systems
}

\author{Sibo Tian$^{1}$, Xiao Liang$^{2}$, Sara Behdad$^{3}$, and Minghui Zheng$^{1,*}$
\thanks{This work was supported by the U.S. National Science Foundation under Grants No. 2527316, No. 2422826, No. 2422640, and No. 2324950.}
\thanks{$^{1}$ Sibo Tian and Minghui Zheng are with the J. Mike Walker '66 Department of Mechanical Engineering, Texas A\&M University, College Station, TX 77843, USA. {\tt\small Emails: \{sibotian, mhzheng\}@tamu.edu.}}
\thanks{$^{2}$ Xiao Liang is with the Zachry Department of Civil and Environmental Engineering, Texas A\&M University, College Station, TX 77843, USA. {\tt\small Email: xliang@tamu.edu.}}
\thanks{$^{3}$ Sara Behdad is with the Engineering School of Sustainable Infrastructure \& Environment, University of Florida, Gainesville, Florida 32611, USA. {\tt\small Email: sarabehdad@ufl.edu.}}
\thanks{$^{*}$ Corresponding Author.}}

\begin{document}

\maketitle
\thispagestyle{empty}
\pagestyle{empty}

\begin{abstract}

Remanufacturing is fundamentally more challenging than traditional manufacturing due to the significant uncertainty, variability, and incompleteness inherent in end-of-life (EoL) products. At the same time, it has become increasingly essential and urgent for facilitating a circular economy, driven by the growing volume of discarded electronic products and the escalating scarcity of critical materials. In this paper, we review the existing literature and examine the key challenges as well as emerging opportunities in intelligent automation for EoL electronics remanufacturing, providing a comprehensive overview of how robotics, control, and artificial intelligence (AI) can jointly enable scalable, safe, and intelligent remanufacturing systems. This paper starts with the definition, scope, and motivation of remanufacturing within the context of a circular economy, highlighting its societal and environmental significance. Then it delves into intelligent automation approaches for disassembly, inspection, sorting, and component reprocessing in this domain, covering advanced methods for multimodal perception, decision-making under uncertainty, flexible planning algorithms, and force-aware manipulation. The paper further reviews several emerging techniques, including large foundation models, human-in-the-loop integration, and digital twins that have the potential to support future research in this area. By integrating these topics, we aim to illustrate how next-generation remanufacturing systems can achieve robust, adaptable, and efficient operation in the face of complex real-world challenges.

\end{abstract}

\section{Introduction}

The rapid advancement of both the economy and technology in recent years has driven a substantial increase in the production and consumption of electronic devices. As a result, the lifecycles of these products have shortened considerably, as consumers frequently upgrade their equipment to keep pace with new features and innovations. Therefore, a growing volume of discarded electronics accumulates, posing serious environmental, resource, and economic challenges, since electronic waste (e-waste) contains both hazardous substances and valuable recoverable materials \cite{ghimire2020wastes}. The situation is further aggravated by insufficient e-waste management infrastructure. In 2022, a record 62 billion kilograms of e-waste were generated globally, of which only 22.3\% was formally documented as being recycled in an environmentally sound manner \cite{balde2024global}. Global e-waste generation is projected to reach 82 billion kilograms by 2030, with only 20\% expected to be properly treated if current practices continue \cite{balde2024global}. Therefore, product and resource recovery have become essential and urgent approaches to mitigate the growing e-waste problem and promote sustainable resource utilization, thereby facilitating a circular economy.

\begin{figure}[t]
    \begin{center}
        \includegraphics[width=0.47\textwidth]{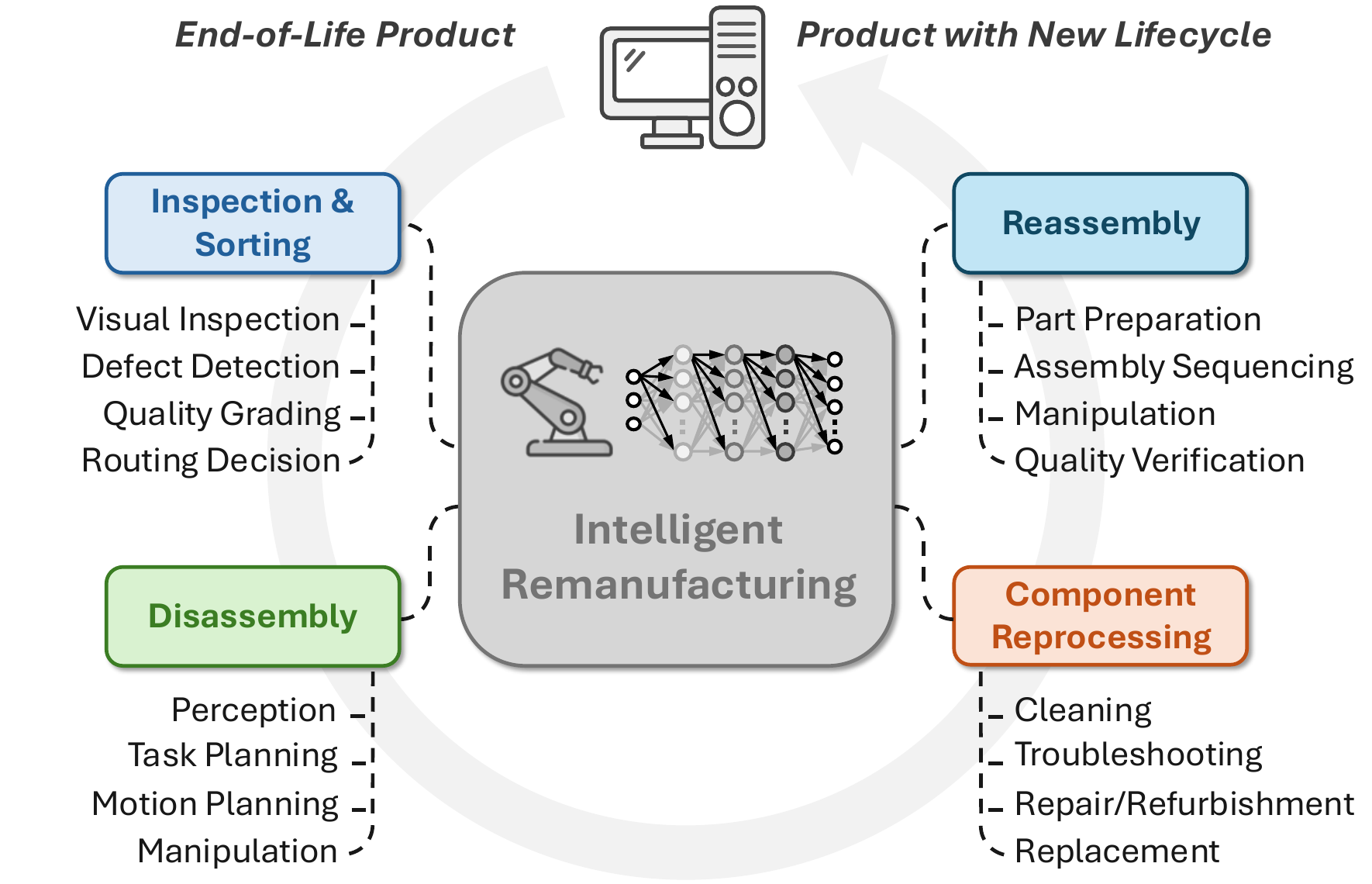}
        \caption{The system-level process of intelligent remanufacturing, supported jointly by advanced robotics, control, and artificial intelligence.
        }
    \label{figure_overview} 
    \end{center}
\vspace{-0.3in}
\end{figure}

Both recycling and remanufacturing are effective strategies for product and resource recovery, helping to limit the disposal of valuable and hazardous materials in landfills. However, they differ in the scope and magnitude of their environmental and economic benefits. Conventional recycling facilities primarily recover raw materials from end-of-life (EoL) products by breaking them down into basic material forms \cite{reck2012challenges}. While this approach reduces landfill waste, it fails to retain the functional and economic value embedded in products and components, preserving very little of their inherent value. On the other hand, as shown in Fig. \ref{figure_overview}, remanufacturing seeks to restore EoL products or components to like-new or better functional performance through systematic processes, including inspection, sorting, disassembly, component reprocessing (e.g., cleaning, repair, and refurbishment), and reassembly \cite{kerin2020smart}. It supports the transition toward a circular economy by extending product lifecycles and minimizing the demand for raw materials.

\begin{figure*}[t]
    \begin{center}
        \includegraphics[width=0.94\textwidth]{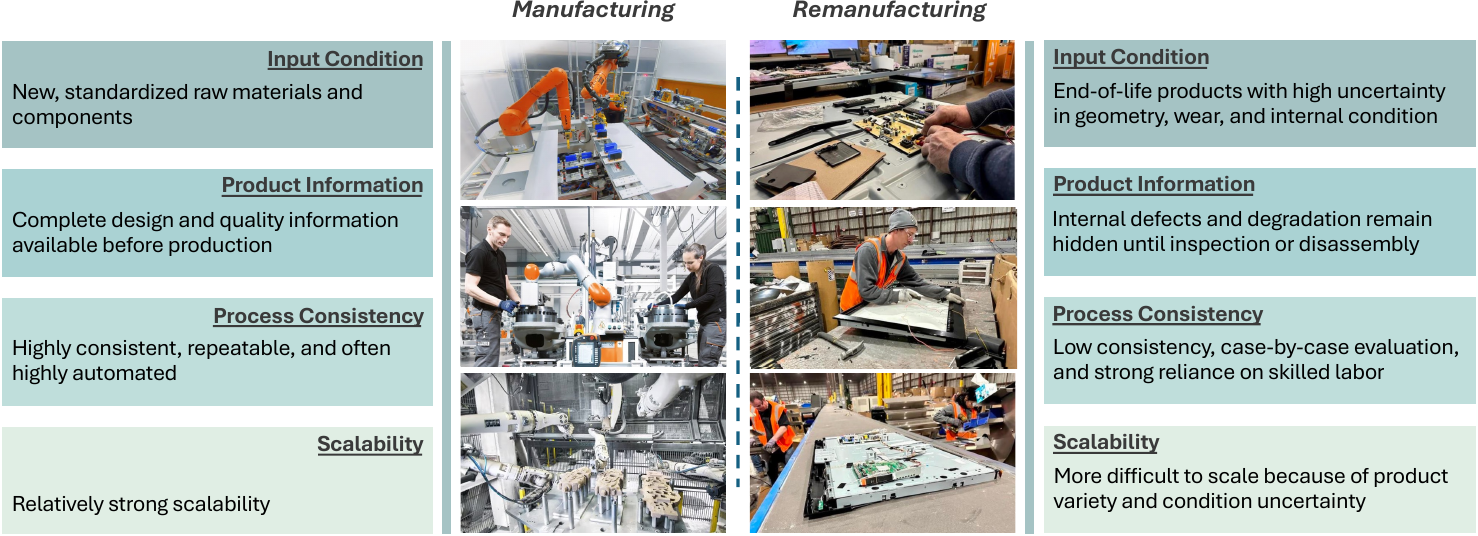}
        \caption{Comparison between manufacturing and remanufacturing. The manufacturing images are sourced from kuka.com, while the remanufacturing images are sourced from Sunnking Sustainable Solutions, Inc. 
        }
    \label{figure_comparison} 
    \end{center}
\vspace{-0.3in}
\end{figure*}

Despite these advantages, remanufacturing is fundamentally complex and challenging, as illustrated by the comparison in Fig. \ref{figure_comparison}.
Traditional manufacturing typically operates on standardized, precisely characterized raw materials and components, following well-established assembly sequences and CAD-guided processes. However, remanufacturing must contend with EoL products that exhibit significant variability and uncertainty in geometry, wear, and internal states \cite{lee2024review}. Specifically, their components may be partially damaged, missing, or tightly coupled due to corrosion, deformation, or prior repair interventions. Moreover, product designs evolve rapidly among different manufacturers and generations, leading to large structural differences even among devices that perform similar functions. As a result, the disassembly sequence, required tools, and manipulation strategies cannot always be predetermined, creating significant challenges for fully automated processing and leaving current remanufacturing operations, particularly disassembly, highly reliant on human labor \cite{xiao2024comprehensive}.

Although humans can handle the unpredictable situations encountered during disassembly, manually disassembling an EoL product often takes longer than assembling it due to tightly interlocked parts, hidden fasteners, and adhesives that complicate the process. The high labor costs and low efficiency associated with manual processes make it difficult to meet the growing demand for e-waste remanufacturing. In addition, many EoL products contain hazardous substances, such as heavy metals, toxic chemicals, and harmful dust, which pose significant risks to human health during manual disassembly. These economic and safety concerns \cite{d2016challenges} together make it difficult to scale current remanufacturing practices.

To tackle these challenges, increasing attention has been directed toward intelligent automation in remanufacturing systems \cite{lu2023state}, offering the potential to enhance operational efficiency, lower labor costs, and reduce human exposure to hazardous materials as well as landfill pollution, thereby boosting both the sustainability and economic viability of these systems. Motivated by this, this paper summarizes the current literature on intelligent automation in remanufacturing, providing an \textbf{\textit{up-to-date}} examination of recent advances, emerging technologies, and practical applications, offering practitioners and researchers a clear understanding of \textbf{\textit{how robotics, control, and artificial intelligence (AI) can together enable scalable, safe, and next-generation remanufacturing systems}}.

\begin{figure}[t]
    \begin{center}
        \includegraphics[width=0.47\textwidth]{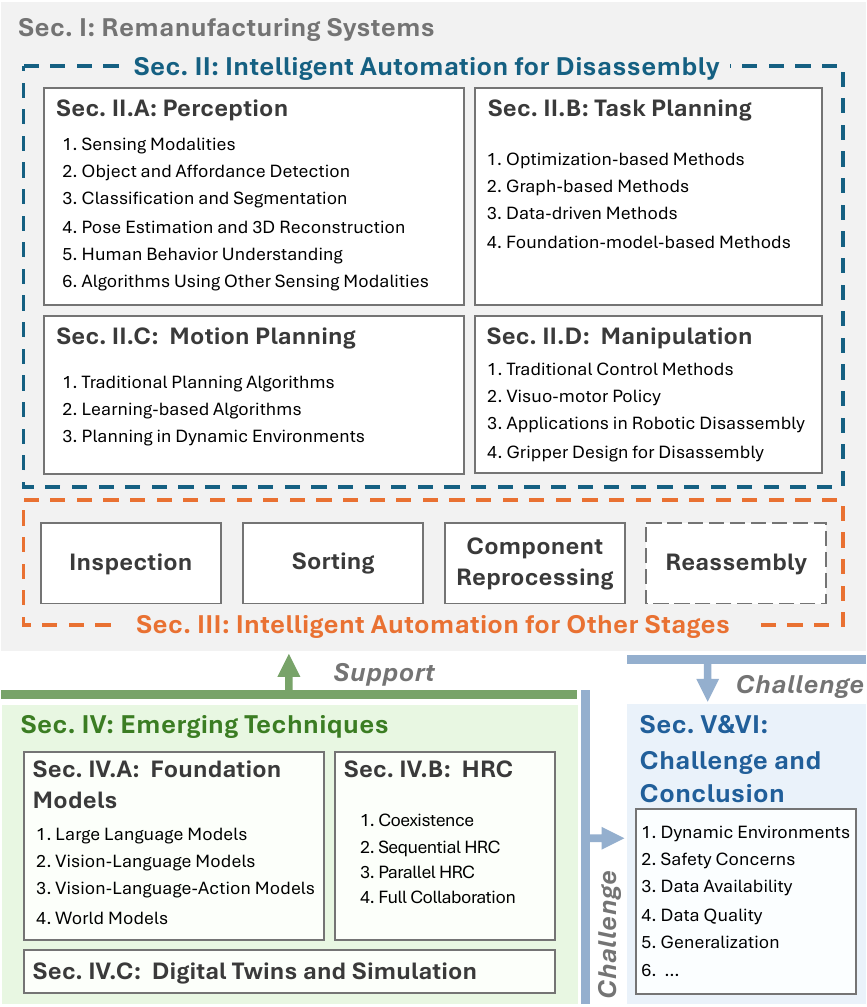}
        \caption{Paper overview. Sections II and III present a system-level review of intelligent automation in remanufacturing. Section IV introduces several emerging techniques to support future research in this domain. Sections V and VI discuss the key challenges and conclude the paper.
        }
    \label{figure_structure} 
    \end{center}
    \vspace{-0.3in}
\end{figure}

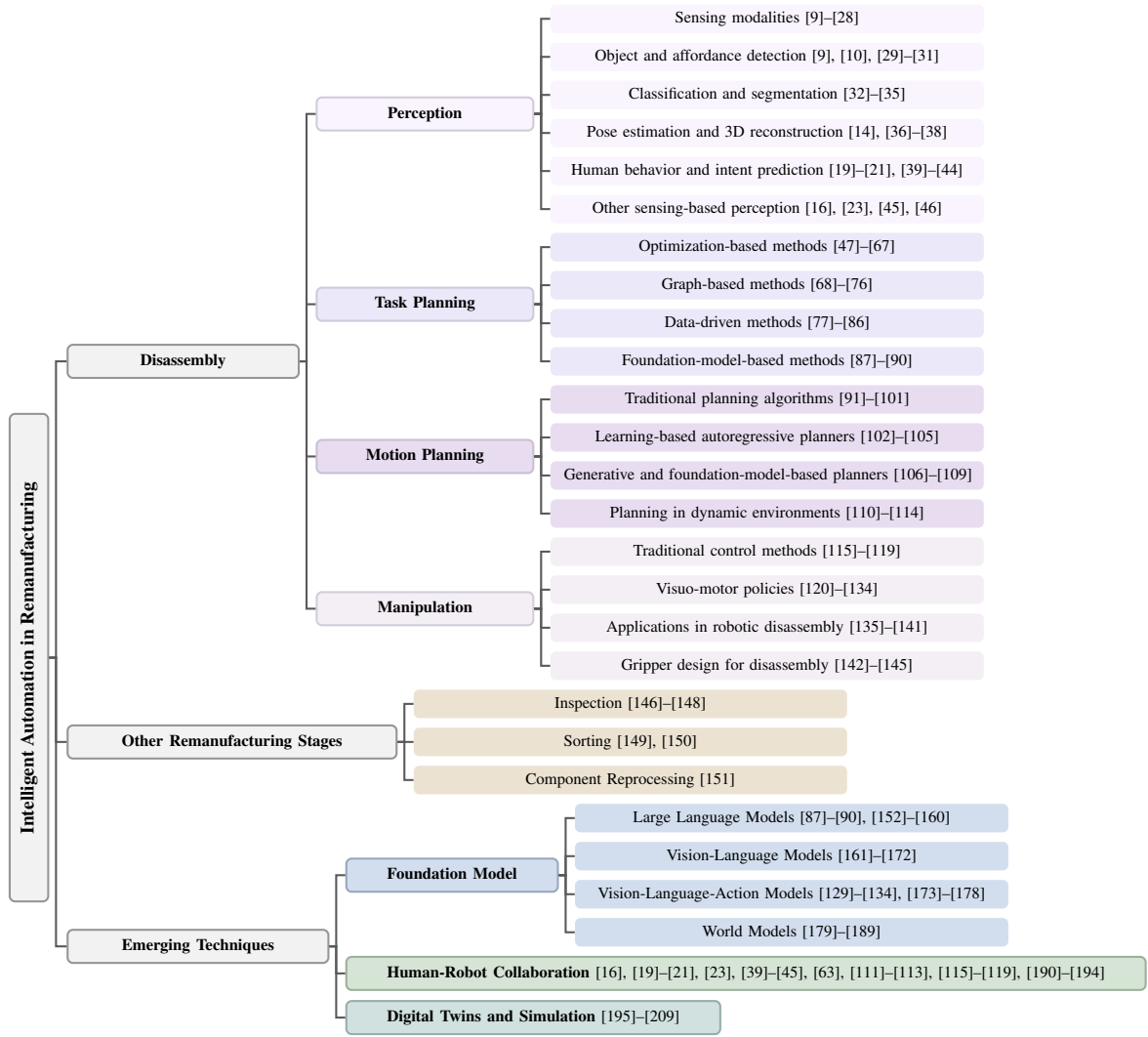
\begin{figure*}[htbp]
    \centering
    \resizebox{0.94\textwidth}{!}
    {
        \definecolor{TitleBorder}{RGB}{149,149,149}
        \definecolor{TitleFill}{RGB}{242,242,242}
        \definecolor{Connector}{RGB}{89,89,89}
        
        \definecolor{C1Border}{RGB}{208,203,217}
        \definecolor{C1Fill}{RGB}{249,244,254}
        
        \definecolor{C2Border}{RGB}{204,203,222}
        \definecolor{C2Fill}{RGB}{234,232,249}
        
        \definecolor{C3Border}{RGB}{200,192,211}
        \definecolor{C3Fill}{RGB}{232,221,240}
        
        \definecolor{C4Border}{RGB}{209,206,217}
        \definecolor{C4Fill}{RGB}{244,239,245}
        
        \definecolor{C5Border}{RGB}{150, 168, 190}
        \definecolor{C5Fill}{RGB}{210, 222, 238}
        
        \definecolor{C6Border}{RGB}{156, 176, 156}
        \definecolor{C6Fill}{RGB}{214, 228, 214}
        
        \definecolor{C7Border}{RGB}{172, 154, 180}
        \definecolor{C7Fill}{RGB}{226, 214, 231}
        
        \definecolor{C8Border}{RGB}{186, 170, 145}
        \definecolor{C8Fill}{RGB}{235, 226, 210}
        
        \definecolor{C9Border}{RGB}{150, 174, 172}
        \definecolor{C9Fill}{RGB}{209, 225, 224}
        
        \tikzset{
          topbox/.style={
            draw=TitleBorder, line width=1.5pt, rounded corners=4pt,
            fill=TitleFill, minimum height=1.0cm,
            inner xsep=50pt, font=\bfseries\Large
          },
          midbox/.style={
            draw=TitleBorder, line width=1.5pt, rounded corners=4pt,
            fill=TitleFill, minimum height=0.86cm, minimum width=6cm,
            inner xsep=40pt, font=\bfseries\large
          },
          group1/.style={
            draw=C1Border, line width=1.5pt, rounded corners=4pt,
            fill=C1Fill, minimum height=0.86cm,
            inner xsep=30pt, font=\bfseries\large, text centered
          },
          group2/.style={
            draw=C2Border, line width=1.5pt, rounded corners=4pt,
            fill=C2Fill, minimum height=0.86cm,
            inner xsep=30pt, font=\bfseries\large, text centered
          },
          group3/.style={
            draw=C3Border, line width=1.5pt, rounded corners=4pt,
            fill=C3Fill, minimum height=0.86cm,
            inner xsep=30pt, font=\bfseries\large, text centered
          },
          group4/.style={
            draw=C4Border, line width=1.5pt, rounded corners=4pt,
            fill=C4Fill, minimum height=0.86cm,
            inner xsep=30pt, font=\bfseries\large, text centered
          },
          group5/.style={
            draw=C5Border, line width=1.5pt, rounded corners=4pt,
            fill=C5Fill, minimum height=0.86cm,
            inner xsep=30pt, font=\bfseries\large, text centered
          },
          group6/.style={
            draw=C6Border, line width=1.5pt, rounded corners=4pt,
            fill=C6Fill, minimum height=0.86cm,
            inner xsep=30pt, font=\large, text centered
          },
          group7/.style={
            draw=C9Border, line width=1.5pt, rounded corners=4pt,
            fill=C9Fill, minimum height=0.86cm,
            inner xsep=30pt, font=\large, text centered
          },
          leaf1/.style={
            draw=none, rounded corners=4pt, fill=C1Fill,
            minimum height=0.74cm, inner xsep=10pt,
            font=\large, text centered
          },
          leaf2/.style={
            draw=none, rounded corners=4pt, fill=C2Fill,
            minimum height=0.74cm, inner xsep=10pt,
            font=\large, text centered
          },
          leaf3/.style={
            draw=none, rounded corners=4pt, fill=C3Fill,
            minimum height=0.74cm, inner xsep=10pt,
            font=\large, text centered
          },
          leaf4/.style={
            draw=none, rounded corners=4pt, fill=C4Fill,
            minimum height=0.74cm, inner xsep=10pt,
            font=\large, text centered
          },
          techleaf/.style={
            draw=none, rounded corners=4pt, fill=C5Fill,
            minimum height=0.74cm, inner xsep=10pt,
            font=\large, text centered
          },
            otherleaf/.style={
            draw=none, rounded corners=4pt, fill=C8Fill,
            minimum height=0.74cm, inner xsep=10pt,
            font=\large, text centered
          },
        }
        \begin{forest}
            forked edges,
            for tree={
                grow=east,
                reversed=true,
                anchor=base west,
                parent anchor=east,
                child anchor=west,
                align=left,
                edge={Connector, line width=1.5pt},
                },
                ver/.style={rotate=90, child anchor=north, parent anchor=south, anchor=center},
            [Intelligent Automation in Remanufacturing, ver, topbox,
                [Disassembly, midbox,
                    [Perception, group1, text width=10em,
                        [{Sensing modalities \cite{mangold2022vision, zhang2023automatic, sun2023novel, song2022deep, suresh2021high, zhao2025precision, diaz2025robotic, song2025adaptive, wang2019flexible, galaiya2025improving, tian2023optimization, tian2025real, zhang2024early, deuerlein2021human, dong2025decoding, wang2024evaluating, hu2021research, calderon2025human, he2025embodied, zhou2024towards}}, leaf1, text width=30em]
                        [{Object and affordance detection \cite{mangold2022vision, zhang2023automatic, puttero2024automatic, myers2015affordance, yin2022object}}, leaf1, text width=30em]
                        [{Classification and segmentation \cite{zhang2026evaluating, liao2025disassembly, noh2025graspsam, wu2023sim2real}}, leaf1, text width=30em]
                        [{Pose estimation and 3D reconstruction \cite{pan2023tax, fan2024integrate, siddiqi2019low, zhao2025precision}}, leaf1, text width=30em]
                        [{Human behavior and intent prediction \cite{tian2024transfusion, tian2025real, eltouny2024tgn, liu2024recurrent, tian2025prediflow, zhang2024early, zhang2025multi, zhang2023unsupervised, tian2023optimization}}, leaf1, text width=30em]
                        [{Other sensing-based perception \cite{song2025adaptive, song2026tatic, dong2025decoding, wang2026bioinspired}}, leaf1, text width=30em]
                    ]
                    [Task Planning, group2, text width=10em,
                        [{Optimization-based methods \cite{go2012genetically, kheder2015disassembly, kucukkoc2020balancing, percoco2013preliminary, liu2018robotic, wang2021discrete, liu2020collaborative, kalayci2013artificial, ccil2022two, luo2016integrated, malik2010performance, yeh2012simplified, kalayci2013particle, guo2017dual, li2013selective, guo2019lexicographic, lee2022task, yin2022mixed, ma2011disassembly, kim2017optimal, zhang2022improved}}, leaf2, text width=30em]
                        [{Graph-based methods \cite{behdad2010disassembly, kuo2010waste, behdad2012disassembly, smith2012disassembly, min2010mechanical, munker2022cad, zhao2014fuzzy, kuo2013waste, guo2015disassembly}}, leaf2, text width=30em]
                        [{Data-driven methods \cite{gao2023data, streibel2024data, zhao2021reinforcement, allagui2023reinforcement, han2023deep, cui2023robotic, amirnia2024context, wang2025deep, xiao2023multi, peng2025dynamic}}, leaf2, text width=30em]
                        [{Foundation-model-based methods \cite{xia2025leveraging, yu2025rescheduling, erdogan2025intent, tong2026gnn}}, leaf2, text width=30em]
                    ]
                    [Motion Planning, group3, text width=10em,
                        [{Traditional planning algorithms \cite{karaman2011sampling, gammell2015batch, janson2015fast, strub2020adaptively, strub2022adaptively, ratliff2009chomp, schulman2014motion, sundaralingam2023curobo, feng2026facto, feng2026drafto, toussaint2015logic}}, leaf3, text width=30em]
                        [{Learning-based autoregressive planners \cite{qureshi2019motion, liu2024kg, soleymanzadeh2025simpnet, soleymanzadeh2026gaide}}, leaf3, text width=30em]
                        [{Generative and foundation-model-based planners \cite{carvalho2023motion, saha2024edmp, tian2025warm, soleymanzadeh2025perfact}}, leaf3, text width=30em]
                        [{Planning in dynamic environments \cite{liu2024hybrid, mainprice2013human, liu2024integrating, park2019planner, yang2025deep}}, leaf3, text width=30em]
                    ]
                    [Manipulation, group4, text width=10em,
                        [{Traditional control methods \cite{chen2014robot, huang2020case, hjorth2023design, zhang2025learning, zang2025augmenting}}, leaf4, text width=30em]
                        [{Visuo-motor policies \cite{chi2025diffusion, ze20243d, ma2024hierarchical, wang2024sparse, ren2024diffusion, cao2025mamba, zhang2025flowpolicy, he2025foar, wu2025tacdiffusion, zitkovich2023rt, kim2024openvla, kim2025fine, black2024pi_0, intelligence2025pi_, intelligence2025pi}}, leaf4, text width=30em]
                        [{Applications in robotic disassembly \cite{kang2025robotic, kang2025task, shukla2025force, lee2025manipforce, liu2025vision, liu2026self, saka2026contact}}, leaf4, text width=30em]
                        [{Gripper design for disassembly \cite{wei2022novel, mountain2024grasping, klas2021kit, zhang2025degrip}}, leaf4, text width=30em]
                    ]
                ]
                [Other Remanufacturing Stages, midbox,
                    [{Inspection \cite{khan2021vision, nwankpa2021achieving, kaiser2026semantic}}, otherleaf, text width=30em]
                    [{Sorting \cite{liu2026raise, mashhadi2017optimal}}, otherleaf, text width=30em]
                    [{Component Reprocessing \cite{zhang2025predictive}}, otherleaf, text width=30em]
                ]
                [Emerging Techniques, midbox,
                    [Foundation Model, group5,
                        [{Large Language Models \cite{brown2020language, touvron2023llama, chowdhery2023palm, minaee2024large, chen2025unleashing, han2024parameter, hu2022lora, ouyang2022training, xia2025leveraging, yu2025rescheduling, erdogan2025intent, tong2026gnn, zhang2025llm}}, techleaf, text width=30em]
                        [{Vision-Language Models \cite{ghosh2024exploring,zhang2024vision,li2025benchmark,radford2021learning,li2022blip,liu2023visual,alayrac2022flamingo,zhou2024transfusion,jiang2024mmad,ueno2025vision,malla2025enhancing,ude2025challenges}}, techleaf, text width=30em]
                        [{Vision-Language-Action Models \cite{ma2024survey,zhang2025pure,zitkovich2023rt,kim2024openvla,kim2025fine,black2024pi_0,intelligence2025pi_,intelligence2025pi,o2024open,yu2025forcevla,huang2025tactile,zhao2026fd}}, techleaf, text width=30em]
                        [{World Models \cite{ding2025understanding,zhang2025step,chen2025egoagent,hong20233d,zhen20243d,du2023video,wu2024ivideogpt,lecun2022path,hafner2019dream,hafner2019learning,li2026learning}}, techleaf, text width=30em]
                    ]
                    [\textbf{Human-Robot Collaboration} \cite{jahanmahin2022human,dhanda2025reviewing,magrini2020human,scimmi2021practical,huang2021experimental,tian2023optimization,tian2025real,tian2024transfusion,eltouny2024tgn,liu2024recurrent,tian2025prediflow,zhang2024early,zhang2025multi,zhang2023unsupervised,song2025adaptive,song2026tatic,dong2025decoding,mainprice2013human,liu2024integrating,park2019planner,chen2014robot,huang2020case,hjorth2023design,zhang2025learning,zang2025augmenting, lee2022task}, group6,
                    ]
                    [\textbf{Digital Twins and Simulation} \cite{jones2020characterising,lu2020digital,wang2019digital,ghorbani2022construction,kerin2023optimising,kerin2023generic,streibel2025integrating,zhao2020sim,mourtzis2020simulation,liu2020real,torne2024reconciling,wagenmaker2024overcoming,park2025dart,fishman2023motion,maddipatla2025vr}, group7,]
                ]
            ]
        \end{forest}
    }
    \caption{Taxonomy of the literature reviewed in this paper on intelligent automation for remanufacturing. The figure organizes the references by remanufacturing stage and technical theme, including disassembly, other remanufacturing stages, and emerging techniques.
    }
    \label{fig:sota-literature}
    \vspace{-0.2in}
\end{figure*}

The remainder of this paper is organized as follows. Sections II and III present a system-level review of intelligent automation in the remanufacturing workflow, with Section II focusing on the disassembly process, which is the most uncertain and challenging stage for automation, and Section III examining intelligent systems applied to other stages of the workflow. Section IV shifts the perspective from a system-level review to emerging techniques and methodologies that support intelligent automation in remanufacturing. Section V summarizes current challenges and highlights future directions and opportunities. Finally, Section VI concludes the paper. The overall structure of this paper is illustrated in Fig. \ref{figure_structure}, and the corresponding references are summarized in Fig. \ref{fig:sota-literature}.

\section{Intelligent Automation for Disassembly}

\begin{figure}[t]
    \begin{center}
        \includegraphics[width=0.47\textwidth]{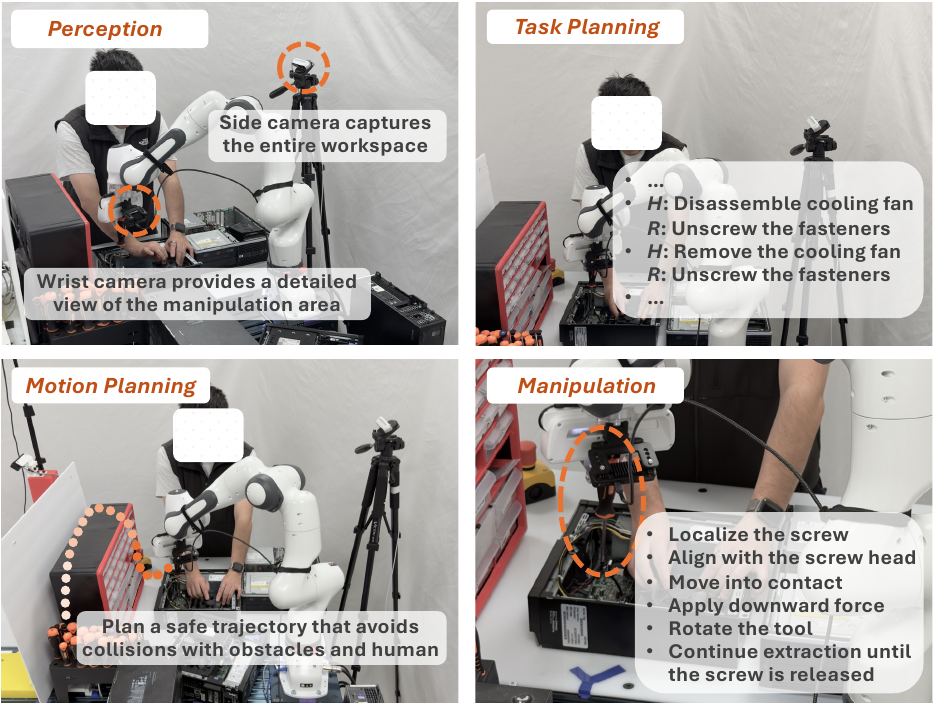}
        \caption{Demonstration of perception, task planning, motion planning, and manipulation in a disassembly example.
        }
    \label{figure_stage} 
    \end{center}
    \vspace{-0.3in}
\end{figure}

Disassembly, as the first operational stage in remanufacturing, aims to systematically decompose EoL electronic products into individual components to facilitate downstream processes such as cleaning, inspection, repair, reassembly, and recycling \cite{lee2024review}. Collected EoL products, such as desktops and smartphones, exhibit significant variability between different brands and models, as well as considerable uncertainty arising from unpredictable usage conditions. Therefore, disassembly exhibits the highest level of complexity among all remanufacturing operations, and is the most challenging stage and critical bottleneck to achieving high levels of automation for sustainable manufacturing. To handle EoL products in a manner comparable to human operators, robotic disassembly systems must integrate intelligent decision-making capabilities with real-time perception and adaptive execution. Typically, an intelligent robotic disassembly system operates in a repetitive loop consisting of perception, task planning, motion planning, and manipulation \cite{lee2024review}. Fig. \ref{figure_stage} showcases an example disassembly task, with each stage illustrated. The remainder of this section examines each of these components in detail.

\begin{figure*}[t]
    \begin{center}
        \includegraphics[width=0.94\textwidth]{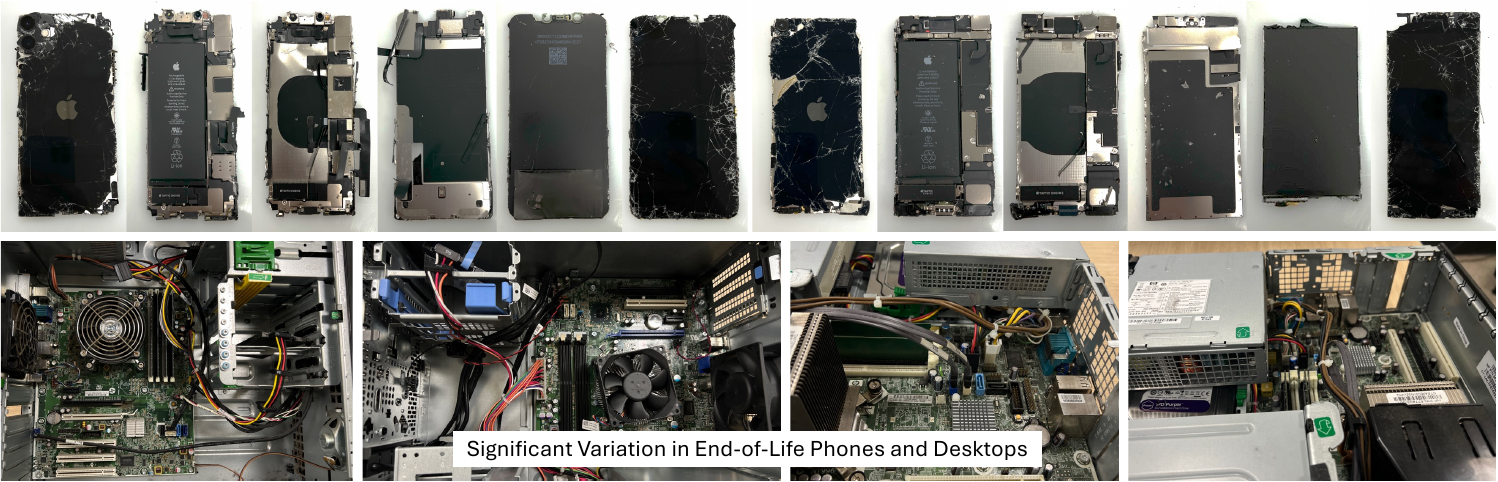}
        \caption{Discarded electronic devices exhibit substantial variation in geometry, wear, and internal states. Densely packed and partially occluded components, as well as tiny and visually similar parts, further complicate robotic perception.
        }
    \label{figure_variation} 
    \end{center}
\vspace{-0.3in}
\end{figure*}

\vspace{-0.1in}
\subsection{Perception}

The perception stage involves acquiring detailed information about the product and its environment. This includes detecting and segmenting individual components, identifying fasteners and connectors, estimating the poses and orientations of parts, and understanding contextual information such as the behaviors and intentions of other collaborators. Accurate perception is critical for subsequent stages, as errors in component detection or pose estimation can propagate through the workflow, potentially causing collisions, failed grasps, or incorrect manipulations. However, perception is particularly challenging in e-waste disassembly, where components are densely packed, partially occluded, and often worn, deformed, or contaminated, as shown in Fig. \ref{figure_variation}. The presence of many small, visually similar parts further complicates reliable detection and pose estimation, while moving human collaborators introduce additional uncertainty into scene understanding and safe state tracking. Therefore, a wide range of sensors and algorithms have been developed and utilized to enhance the perception capabilities of intelligent disassembly systems.

\subsubsection{\textbf{Sensing Modalities}}

Advanced sensing modalities such as RGB \cite{mangold2022vision, zhang2023automatic}, RGB-D cameras \cite{sun2023novel, song2022deep}, 3D image scanners \cite{suresh2021high, zhao2025precision}, force/torque \cite{ diaz2025robotic, song2025adaptive}, and tactile \cite{wang2019flexible, galaiya2025improving} sensors, are often employed in current robotic automation systems. In cases where humans are present as collaborators, the robot must also perceive and interpret human actions and intentions through sensing modalities such as 3D motion capture systems \cite{tian2023optimization, tian2025real, zhang2024early}, voice \cite{deuerlein2021human}, electromyography (EMG) \cite{dong2025decoding, wang2024evaluating}, gestures \cite{hu2021research}, and gaze-tracking \cite{calderon2025human} sensors. These modalities are often combined together to form a multimodal perception system \cite{he2025embodied, zhou2024towards}, enabling robots to integrate information from multiple sources and respond safely and effectively to both objects and environments in real time.

\subsubsection{\textbf{Object and Affordance Detection}}

From an algorithmic perspective, computer vision technologies have been extensively explored in this domain. For object and affordance detection, the algorithms need to detect and localize individual components, fasteners, and tools in cluttered or partially occluded environments. They also need to go beyond recognizing what an object is and predict how it can be interacted with. Mangold et al. \cite{mangold2022vision} employ YOLOv5 to detect and classify six different types of screw heads. Similarly, Zhang et al. \cite{zhang2023automatic} leverage YOLOv4 \cite{bochkovskiy2020yolov4} to detect and localize screws in EoL desktops. They further employ EfficientNetV2 \cite{tan2021efficientnetv2} to classify the detected screws and recommend the corresponding tools. In addition to screws, Puttero et al. \cite{puttero2024automatic} detect other components, such as wires and buttons, on EoL electronic boards. For affordance detection, Myers et al. \cite{myers2015affordance} use local shape and geometric primitives to predict object affordances using a linear support vector machine (SVM). Yin et al. \cite{yin2022object} propose a boundary-preserving network to generate affordance masks with improved boundary quality for robotic manipulation.

\subsubsection{\textbf{Classification and Segmentation}}

As for classification and segmentation, the algorithms aim to assign semantic labels to different components and separate them from the background or surrounding parts. This capability is essential for identifying specific components and understanding their spatial extents \cite{zhang2026evaluating}. For instance, Liao et al. \cite{liao2025disassembly} detect component sizes through classification and segmentation, and then evaluate a disassembly score to assess whether a given robot is capable of performing specific disassembly tasks. In addition, Segment Anything Model (SAM) \cite{kirillov2023segment} is adapted in \cite{noh2025graspsam} to determine object grasp positions, integrating object segmentation and grasp prediction into a unified framework. In addition to 2D images, a 3D point cloud segmentation network \cite{wu2023sim2real} is trained on synthetic data and fine-tuned on real-world point clouds to achieve sim-to-real transfer.

\subsubsection{\textbf{Pose Estimation and 3D Reconstruction}}

For pose estimation and 3D reconstruction, the goal is to estimate the position and orientation of objects and reconstruct their 3D geometry from 2D images, depth maps, or point clouds, enabling precise planning and manipulation. Pan et al. \cite{pan2023tax} estimate the relative pose between two objects for a given manipulation task, such as the pose of a mug relative to a mug rack, to guide the robot in manipulating the objects into the desired configuration. Fan et al. \cite{fan2024integrate} focus on estimating 3D hand-object interaction poses to mitigate occlusions in 2D images that occur during close human-robot collaborative disassembly (HRCD). A low-cost method is proposed in \cite{siddiqi2019low} to reconstruct a 3D model for disassembly automation from multiple 2D images. Zhao et al. \cite{zhao2025precision} employ a seven-color illumination strategy and leverage the spectral response characteristics of a color camera for different wavelengths, effectively mitigating 3D reconstruction errors caused by overexposure on highly reflective surfaces in EoL desktops.

\subsubsection{\textbf{Human Behavior Interpretation and Prediction}}

When one or more human collaborators are present, the robot needs to infer human behavior and intent in order to coordinate safely and efficiently with them. Tian et al. \cite{tian2024transfusion} propose a generative model-based motion prediction algorithm aimed at robust and stochastic long-term human motion prediction. They further leverage knowledge distillation to enhance the algorithm’s responsiveness and achieve real-time prediction performance in HRCD \cite{tian2025real}. Eltouny et al. \cite{eltouny2024tgn} use deep ensembles to predict human motion with uncertainty estimation. Liu et al. \cite{liu2024recurrent} combine a recurrent neural network with an unscented Kalman filter to predict human arm motion in disassembly, explicitly incorporating a dynamic model linking forces and motion to improve prediction performance. In addition, a coarse-to-fine prediction strategy leveraging contextual information is proposed in \cite{tian2025prediflow}. As for intent prediction, Zhang et al. \cite{zhang2024early} leverage Transformer and Long Short-Term Memory (LSTM) networks to recognize human intent before the motion is completed. They further propose a unified network that simultaneously predicts human intent and future motion trajectories using multi-task learning techniques \cite{zhang2025multi}. An unsupervised learning method is proposed in \cite{zhang2023unsupervised}, allowing robots to autonomously learn to predict human intent for hard disk disassembly. Moreover, to enhance interpretability, Tian et al. \cite{tian2023optimization} assume that humans are rational agents and utilize inverse optimal control to recover the underlying cost function that governs human motion, which can then be used to predict motion by solving the forward optimal control problem.

\subsubsection{\textbf{Algorithms Using Other Sensing Modalities}}

In addition to computer vision technologies, various methodologies from different research domains are employed to enhance robotic perception. Song et al. \cite{song2025adaptive} propose an algorithm that enables a robotic manipulator to detect human intervention and correct its motion based on torque sensor feedback. The authors further employ torque-based contact force estimation together with a task-aware Temporal Convolutional Network (TCN) to jointly infer discrete task-level intent and continuous motion-level parameters \cite{song2026tatic}. Dong et al. \cite{dong2025decoding} design a soft sensor attached to the human face to decode silent speech from muscular biopotential signals for applications in HRCD. Wang et al. \cite{wang2026bioinspired} develop a bioinspired textured sensor array with a circular grid arrangement, significantly reducing the robot’s response time for object classification.

Overall, advances in sensing and perception algorithms significantly enhance robotic understanding of complex disassembly environments and human behavior, providing a strong foundation for subsequent decision-making and manipulation.

\vspace{-0.1in}
\subsection{Task Planning}

Task planning is one of the most important decision-making layers in intelligent automation systems for disassembly. In general, it determines which tasks are required to break down EoL products, specifies their execution order under structural and accessibility constraints, and, in collaborative scenarios, determines task allocation while considering both safety and operational efficiency. 
However, unlike assembly, which is typically deterministic and supported by complete product information, disassembly task planning is significantly more challenging. It must address uncertainties arising from the unknown conditions of EoL products during the disassembly process. In addition, product documentation and detailed CAD models are often unavailable, further complicating the planning process. As a result, disassembly planning represents a highly complex combinatorial optimization problem subject to real-time changes, motivating the development of computational methods for generating feasible and efficient disassembly sequences.

\subsubsection{\textbf{Optimization-based Methods}}

Classical optimization algorithms have been widely applied in disassembly task planning, such as genetic algorithms \cite{go2012genetically, kheder2015disassembly, kucukkoc2020balancing}, artificial bee colony \cite{percoco2013preliminary, liu2018robotic, wang2021discrete, liu2020collaborative, kalayci2013artificial, ccil2022two}, ant colony \cite{luo2016integrated, malik2010performance}, particle swarm \cite{yeh2012simplified, kalayci2013particle}, scatter search \cite{guo2017dual, li2013selective, guo2019lexicographic}, and linear programming \cite{lee2022task, yin2022mixed, ma2011disassembly, kim2017optimal, zhang2022improved}. In general, the objective function of these optimization problems may consist of a single term or a combination of several terms, including total disassembly time, associated costs, environmental impact, workloads, and safety. The algorithms are typically designed either to minimize computation time or to improve the quality of optimization outcomes. Although these methods have demonstrated promising results, they generally assume complete information regarding the product structure, component conditions, operational constraints, and the time required for processing each component, which limits the applicability of these methods in real-world scenarios.

\subsubsection{\textbf{Graph-based Methods}}

Graph-based modeling represents another approach to disassembly task planning. Directed or undirected graphs, AND/OR graphs, and Petri Net (PN) modeling are three commonly used representations in this domain. A graph representation \cite{behdad2010disassembly, kuo2010waste, behdad2012disassembly, smith2012disassembly}, either directed or undirected, typically uses nodes to represent product components or subassemblies and edges to capture their connections or precedence constraints. While these methods effectively represent relationships among components, they generally do not explicitly encode alternative sequences and are less flexible when multiple disassembly routes or parallel operations are possible. AND/OR graphs \cite{min2010mechanical, munker2022cad}, are specifically designed to represent multiple feasible disassembly strategies. In these graphs, an AND relationship indicates that all child nodes must be completed together, whereas an OR relationship signifies that a node can be decomposed through one of several alternative methods. PN modeling \cite{zhao2014fuzzy, kuo2013waste, guo2015disassembly} is a mathematical and graphical tool used to model and analyze discrete-event systems, particularly those with concurrent, parallel, or asynchronous processes. Although these graph-based representations clearly visualize the disassembly process and can be combined with search algorithms to generate disassembly plans, they become increasingly difficult to manage for complex products as graph size and complexity grow rapidly.

\subsubsection{\textbf{Data-driven Methods}}

In real-world scenarios without privileged information, it is difficult to predetermine a feasible disassembly sequence for EoL products due to inherent uncertainties, further limiting the practical deployment of the previously discussed methods. To address these limitations, data-driven and learning-based methods for dynamic disassembly sequencing and task allocation under uncertainty have been increasingly explored. Gao et al. \cite{gao2023data} quantify random and fuzzy assessment data associated with uncertainty and leverage a data-driven model to predict the turning time of disassemblability for a given certainty degree. The predicted values are then used to determine the optimal disassembly sequence. Streibel et al. \cite{streibel2024data} use real-time observations, simulation-generated data, and a historical experience database to develop a data-driven method for online replanning during disassembly when new information deviates from prior assumptions. Reinforcement learning (RL) has been widely used for adaptive online disassembly planning in uncertain scenarios \cite{zhao2021reinforcement, allagui2023reinforcement, han2023deep, cui2023robotic, amirnia2024context, wang2025deep, xiao2023multi, peng2025dynamic}. In general, these methods iteratively learn optimal policies that maximize a reward function through trial and error, selecting the next best disassembly task or the most effective task allocation in HRCD settings based on the observed state.

\begin{figure*}[t]
    \begin{center}
        \includegraphics[width=0.94\textwidth]{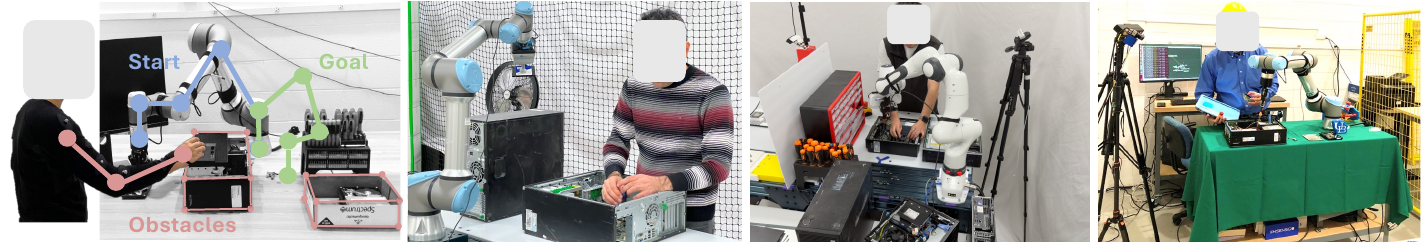}
        \caption{The workspace of e-waste disassembly is typically cluttered, unstructured, and dynamic, making real-time motion planning highly challenging.
        }
    \label{figure_hrc_planning} 
    \end{center}
\vspace{-0.3in}
\end{figure*}

\subsubsection{\textbf{Foundation-model-based Methods}}

With recent advances in large language models (LLMs), several studies have leveraged their semantic understanding and reasoning capabilities to enhance disassembly task planning. Xia et al. \cite{xia2025leveraging} propose using a fine-tuned LLM to quantitatively evaluate disassembly sequences generated by a Bayesian network. High-performing sequences identified by the LLM are then fed back to update the Bayesian network, thereby improving overall planning performance. Yu et al. \cite{yu2025rescheduling} address the online task rescheduling problem in HRCD settings by leveraging a multimodal LLM for disassembly scene understanding. The LLM generates a semantic scene graph, which is then used to query an affordance knowledge graph, matching similar disassembly environments and providing disassembly step information. Erdogan et al. \cite{erdogan2025intent} generate a structured textual representation of the scene from RGB-D images and query a LLM to produce multiple candidate disassembly sequences. They then employ a second LLM to verify the logical ordering and precedence of these sequences. Tong et al. \cite{tong2026gnn} use an LLM to improve task planning in multi-human multi-robot collaboration by providing high-level semantic understanding of tasks, agents, operating procedures through text-attributed knowledge representations. These works underscore the growing importance of large foundation models in disassembly sequencing and task allocation. With their strong semantic reasoning capabilities and rich prior knowledge acquired through large-scale pretraining, LLMs represent a promising solution for dynamic and flexible disassembly task planning in complex, single- or multi-agent disassembly systems.

\vspace{-0.1in}
\subsection{Motion Planning}

Motion planning is another important decision-making layer in intelligent disassembly systems. While task planning determines which operations to perform and in what order, motion planning focuses on generating feasible, safe trajectories for physical execution. It is usually defined as computing a trajectory from an initial configuration to a goal configuration, or to a target end-effector pose, while satisfying a set of constraints, including kinematic and dynamic limitations as well as safety requirements \cite{soleymanzadeh2026towards}. In disassembly scenarios, motion planning is particularly challenging due to cluttered, confined, and dynamic workspaces, as shown in Fig. \ref{figure_hrc_planning}. In such settings, motion planners must be effective and support frequent replanning to ensure safe operation, especially in multi-agent collaborative environments.

\subsubsection{\textbf{Traditional Planning Algorithms}}

Sampling-based and optimization-based algorithms are two well-established approaches for robot motion planning. Sampling-based methods, such as RRT* \cite{karaman2011sampling}, which is asymptotically optimal given sufficient planning time, BIT* \cite{gammell2015batch} and FMT* \cite{janson2015fast}, which improve planning efficiency, AIT* \cite{strub2020adaptively}, which further enhances heuristic guidance through problem-specific heuristics, and EIT* \cite{strub2022adaptively}, which combines cost and effort heuristics for robust performance, are well suited to high-dimensional motion planning problems. However, they often struggle with task-constrained planning and real-time responsiveness in complex environments. Optimization-based algorithms, such as CHOMP \cite{ratliff2009chomp}, TrajOpt \cite{schulman2014motion}, cuRobo \cite{sundaralingam2023curobo}, FACTO \cite{feng2026facto}, and DRAFTO \cite{feng2026drafto}, treat robot motion planning as a constrained trajectory optimization problem, where the objective typically integrates trajectory smoothness, collision avoidance, and task-specific costs. By iteratively refining the trajectory, these methods can directly generate smooth, high-quality motions that satisfy multiple constraints without post-processing. However, optimization algorithms are highly sensitive to the initial guess and are prone to becoming trapped in local minima. When the initial trajectory seed is poor, the path may penetrate obstacles deeply, making recovery and convergence to a feasible solution difficult. Usually, if a linearly interpolated seed fails, a sampling-based planner is used to provide a better initial guess, trading planning time for improved feasibility. \cite{toussaint2015logic}.

\subsubsection{\textbf{Learning-based Algorithms}}

Recently, deep learning methods such as imitation learning have been increasingly used to train surrogate models that mimic traditional planners. Their training data are typically generated by optimal planners such as RRT* \cite{karaman2011sampling} or cuRobo \cite{sundaralingam2023curobo}. An MLP-based neural motion planner is proposed in \cite{qureshi2019motion} to predict the next robot joint configuration toward the goal. It trains an autoencoder to compress the environment point cloud into a latent space and a planner network that takes the current and goal configurations together with the latent environment representation as input. Liu et al. \cite{liu2024kg} propose a bidirectional neural motion planner that solves the problem simultaneously from the start and goal. It checks collisions along the interpolated path between the two planning processes and uses a graph neural network (GNN), rather than an MLP, to predict the next robot configuration and capture spatial dependencies between the robot and nearby obstacles. Similarly, Soleymanzadeh et al. \cite{soleymanzadeh2025simpnet} encode the robot's kinematic structure as a graph and use cross-attention to integrate workspace information such as obstacle geometry and position. Dropout is applied during inference to generate alternative paths if a previous attempt results in a collision. Their follow-up work \cite{soleymanzadeh2026gaide} uses a point cloud encoder to downsample robot and obstacle points, which are then incorporated into a graph to encode spatial information. Although these methods speed up planning with neural networks, they still require rollout to generate a full trajectory, since they only predict the next waypoint.

Generative models have also been explored for stochastic motion planning. Rather than predicting the next waypoint, these methods operate directly on the full trajectory. Carvalho et al. \cite{carvalho2023motion} train a diffusion-based generative model over valid trajectories for robot manipulator motion planning. During inference, a customized cost function guides denoising toward lower-cost trajectory regions, gradually producing motions that satisfy constraints such as smoothness and collision avoidance. Saha et al. \cite{saha2024edmp} extend this work by using an ensemble of cost functions with different weights and solving planning problems in parallel, improving the success rate across scenarios. Instead of end-to-end planning, Tian et al. \cite{tian2025warm} propose a point cloud-conditioned flow matching model for learning robot motion in cluttered environments without privileged information. The predicted plans are then used as initial seeds for batched trajectory optimization, achieving high success rates and real-time performance. Moreover, Soleymanzadeh et al. \cite{soleymanzadeh2025perfact} propose an LLM-powered method to generate diverse, semantically feasible workspaces for large-scale planning data collection. The authors also introduce a neural motion planner based on a fusion action-chunking transformer for improved planning-signal encoding and multimodal attention \cite{soleymanzadeh2025perfact}.

\subsubsection{\textbf{Planning in Dynamic Environments}}

Despite their strong performance, the methods discussed above mainly focus on static environments. In disassembly, however, the workspace is often dynamic, as component removal, tool interactions, and human involvement continuously change the workspace configuration. To remain effective in such settings, planning algorithms must either account for moving obstacles explicitly or be fast enough to support frequent replanning; in some cases, both are needed. Liu et al. \cite{liu2024hybrid} propose a method that enables a robot to alter its joint configuration with minimal reconfiguration effort while maintaining the end-effector position when a potential collision with a human operator is detected. Mainprice et al. \cite{mainprice2013human} forecast human workspace occupancy by computing the swept volume of the predicted trajectory. The robot then plans its motion by minimizing penetration cost while interleaving planning and execution. Liu et al. \cite{liu2024integrating} address manipulator motion planning in HRCD by incorporating uncertainty-aware human motion prediction into a graph-based planner, enabling proactive responses to human movement and reducing the need for frequent replanning. Similarly, Park et al. \cite{park2019planner} propose an intention-aware online planner that models predicted human motion as a Gaussian distribution and computes tight upper bounds on collision probability for safe robot motion planning. Instead of relying on motion prediction, Yang et al. \cite{yang2025deep} improve planning robustness in dynamic environments by training a visuo-motor neural motion policy on a large-scale dataset and incorporating a dynamic closest-point Riemannian motion policies module at inference time for local adjustments to avoid collisions.

Overall, learning-based planners retain the robustness of classical methods while significantly improving planning speed. By integrating predictive models, they can also anticipate potential collisions, enabling safer and more proactive trajectory generation in dynamic disassembly environments.

\begin{figure*}[t]
    \begin{center}
        \includegraphics[width=0.94\textwidth]{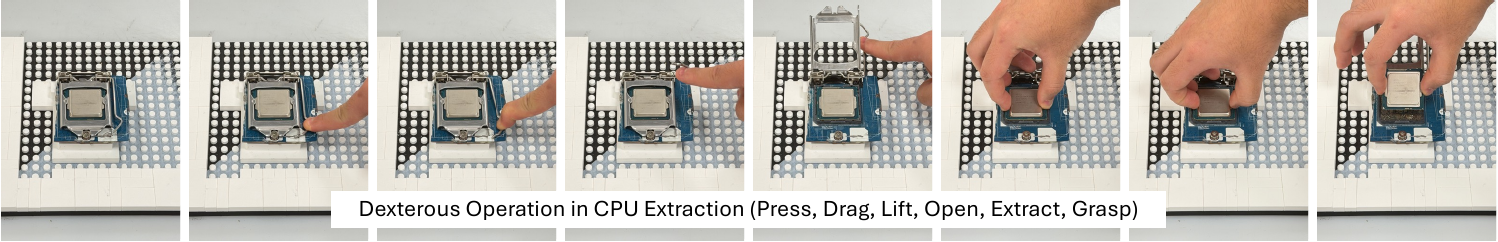}
        \caption{Complex and dexterous disassembly operations in CPU extraction.
        }
    \label{dexterous_op} 
    \end{center}
    \vspace{-0.3in}
\end{figure*}

\vspace{-0.1in}
\subsection{Manipulation}

After motion planning brings the robot to the task area, it must use sophisticated manipulation skills to execute the planned operations. In disassembly, manipulation is especially challenging due to confined workspaces and the need to handle small or tightly coupled components. These tasks often involve dexterous operations, as shown in Fig. \ref{dexterous_op}, and complex interactions such as grasping, unscrewing, cutting, lifting, and separating hard-to-access or tightly fastened parts. In addition, different operations may require different tool heads, actuation forces, and levels of precision, further increasing the complexity of disassembly manipulation. Under these conditions, advanced robot control algorithms are essential for ensuring safe and effective interaction with EoL components.

\subsubsection{\textbf{Traditional Control Methods}}

Impedance control \cite{chen2014robot, huang2020case, hjorth2023design, zhang2025learning, zang2025augmenting}, which enables robots to regulate contact interactions by adjusting virtual stiffness, damping, and inertia, is widely used in robotic disassembly tasks involving contact forces. Chen et al. \cite{chen2014robot} employ an impedance-controlled robot in an HRCD workstation for unscrewing, showing that low joint impedance and Cartesian stiffness adjustment can compensate for positioning errors and tolerate uncertainties during manual positioning. Huang et al. \cite{huang2020case} investigate HRCD of press-fitted components using Cartesian impedance control. The robot uses active compliance to grasp the extracted part and follow its motion while the human operator applies force during separation. Hjorth et al. \cite{hjorth2023design} propose an energy-aware Cartesian impedance controller for unplanned interactions and contact losses in collaborative disassembly. By combining energy scaling, damping injection, and an augmented energy tank, the controller limits the instantaneous power delivered from the controller to the robot during contact loss, thereby ensuring safe interactions. Zhang et al. \cite{zhang2025learning} propose an EMG-driven variable impedance controller that maps human muscle EMG signals to robot joint stiffness, enabling human-like compliant behavior and online stiffness adaptation during contact-rich cable disassembly. Zang et al. \cite{zang2025augmenting} combine Cartesian impedance control with RL to realize a twist-pulling strategy, allowing robots to generalize disassembly skills across varying cap-shaft configurations while handling task uncertainties.

\subsubsection{\textbf{Visuo-motor Policy}}

Despite being effective for regulating physical interactions in robotic disassembly, these traditional methods usually require task-specific controller design and careful parameter tuning. On the other hand, visuo-motor control methods directly couple visual perception with robot actions, offering a promising alternative for manipulation. With rapid advances in generative models, research on robot manipulation has expanded rapidly in recent years.

Diffusion Policy \cite{chi2025diffusion} formulates visuo-motor policy learning as a conditional denoising process over action sequences, allowing it to handle multimodal actions and high-dimensional action spaces. It takes image observation sequences as input, uses a time-series transformer to generate short-horizon action trajectories, and controls the robot in a receding-horizon manner for closed-loop replanning with temporal consistency. Follow-up works have improved diffusion policy in many different ways. 3D diffusion policy (DP3) \cite{ze20243d} replaces 2D image conditioning with compact 3D point-cloud representations, improving training data efficiency and model performance. Hierarchical diffusion policy (HDP) \cite{ma2024hierarchical} introduces a hierarchical and kinematics-aware formulation that separates high-level next-best-pose planning from low-level diffusion-based trajectory generation. Sparse diffusion policy (SDP) \cite{wang2024sparse} incorporates a mixture-of-experts architecture to support multitask, continual, and transferable robot learning with far fewer active parameters. Diffusion Policy Policy Optimization (DPPO) \cite{ren2024diffusion} treats the denoising process as a Markov Decision Process (MDP) and incorporates RL to fine-tune the entire chain of diffusion MDPs using policy gradient. Mamba Policy \cite{cao2025mamba} and FlowPolicy \cite{zhang2025flowpolicy} focus on efficiency, reducing computational cost and inference latency while maintaining task performance. FoAR \cite{he2025foar} and TacDiffusion \cite{wu2025tacdiffusion} extend diffusion policies to contact-rich manipulation by incorporating force/torque or tactile information.

In addition to the diffusion policy family, Vision-Language-Action (VLA) models represent another visuo-motor control approach. RT-2 \cite{zitkovich2023rt} shows that a large vision-language model (VLM) can be co-fine-tuned on web-scale vision-language data and robot trajectories by representing robot actions as text tokens, yielding strong semantic generalization and improved performance on novel objects and instructions. OpenVLA \cite{kim2024openvla} advances this paradigm toward open and scalable robot learning, improving performance while remaining practical for efficient fine-tuning and deployment. OpenVLA-OFT \cite{kim2025fine} proposes an optimized fine-tuning recipe with parallel decoding, action chunking, continuous action representations, and an L1 regression objective, substantially improving control speed and task success rate over standard OpenVLA fine-tuning. Additionally, $\pi_0$ \cite{black2024pi_0} combines a high-level VLM with a small flow matching-based action head for continuous action generation. $\pi_{0.5}$ \cite{intelligence2025pi_} emphasizes co-training on heterogeneous data sources to better capture semantic context, task structure, and cross-robot skill transfer, improving open-world generalization. $\pi_{0.6}^*$ \cite{intelligence2025pi} further advances the VLA framework by incorporating RL from real-world deployment through advantage-conditioned policy training, enabling continuous improvement through interaction and correction.

\subsubsection{\textbf{Applications in Robotic Disassembly}}

While the above work focuses on general manipulation skills, several studies \cite{kang2025robotic, kang2025task, shukla2025force, lee2025manipforce, liu2025vision, saka2026contact} have already tested these models directly on robotic disassembly tasks. Kang et al. \cite{kang2025robotic} apply a vision-force diffusion policy to compliant object prying in battery removal. Cross-attention is utilized to preserve force conditional information, and the resulting features are used to guide policy generation. The authors further incorporate language guidance to provide task context for multi-task disassembly \cite{kang2025task}. Shukla et al. \cite{shukla2025force} train a diffusion policy from human demonstrations and incorporate real-time force feedback for dual-arm compliant sheet separation. Lee et al. \cite{lee2025manipforce} focus on contact-rich manipulation, such as battery disassembly, and train a transformer-based diffusion policy that encodes asynchronous RGB and force/torque signals using frequency- and modality-aware embeddings, fusing them via bi-directional cross-attention for robust multimodal integration. Besides diffusion policies, Liu et al. \cite{liu2025vision} test VLA models on a selective desktop disassembly task and propose a rule-based strategy-integrated framework for precision-required contact-rich disassembly. The authors further extend their work to formulate an agentic VLA, comprising a VLA-planner, a skills library, and a VLA-corrector, to enable autonomous decision-making and error recovery in complex EoL desktop disassembly tasks \cite{liu2026self}. Saka et al. \cite{saka2026contact} systematically investigate the role of tactile sensing in robotic disassembly, suggesting that compact and structured force encoding is more effective than high-dimensional tactile imagery in contact-rich disassembly tasks.

\subsubsection{\textbf{Gripper Design for Disassembly}}

Besides control algorithms, gripper design also strongly affects robotic disassembly. Because disassembly often requires dexterous actions such as prying, twisting, handling irregular parts, unscrewing, or separating press-fitted components, traditional two-finger grippers are often inadequate. This has driven the development of adaptive, multi-fingered, and multifunctional grippers for diverse disassembly tasks. Wei et al. \cite{wei2022novel} design a lightweight, cable-driven, and cost-effective robotic hand for dexterous manipulation. Mountain et al. \cite{mountain2024grasping} further develop an iterative learning control algorithm for this 3D-printed robotic hand, enhancing grasping performance through feedforward adaptation automatically. Klas et al. \cite{klas2021kit} design a compact 7-degree-of-freedom (DoF) gripper equipped with force/pressure sensing and vision capabilities, capable of performing a wide range of actions required for the e-waste disassembly, including picking, dropping, levering, tool changing, flipping, shaking, unscrewing, pushing, cutting, and in-hand object repositioning. Zhang et al. \cite{zhang2025degrip} design a compact cable-driven gripper for disassembly in confined spaces, specifically tailored for handling tightly packed components inside EoL desktops.

Together, these studies highlight that successful robotic disassembly depends on both advanced manipulation algorithms and gripper designs that provide sufficient dexterity, adaptability, and multimodal sensing capabilities. Progress in this area has opened promising new opportunities for intelligent disassembly automation.

\section{Intelligent Automation for Other Remanufacturing Processes}

Beyond disassembly and reassembly, which largely resembles assembly automation in traditional manufacturing processes, intelligent automation has also been widely explored in other remanufacturing stages, including inspection \cite{khan2021vision, nwankpa2021achieving, kaiser2026semantic}, sorting \cite{liu2026raise, mashhadi2017optimal}, and component reprocessing \cite{zhang2025predictive}.

Khan et al. \cite{khan2021vision} develop a vision-guided ultrasonic inspection system for non-destructive product testing, where a robot-mounted RGB camera first reconstructs the part geometry and the resulting 3D model is then used to generate customized ultrasonic tool paths for flexible inspection of complex remanufactured components without relying on accurate CAD models. Nwankpa et al. \cite{nwankpa2021achieving} propose a deep learning-based vision inspection system for remanufacturing, using a ResNet-based convolutional neural network (CNN) to classify multiple surface defects. Kaiser et al. \cite{kaiser2026semantic} propose a semantic 3D product modeling approach for automated remanufacturing inspection, integrating RGB images and 3D point clouds to build semantic product models that encode component and defect information for adaptive visual inspection without prior CAD models. Liu et al. \cite{liu2026raise} propose a robot-assisted selective disassembly and sorting system for EoL phones that integrates adaptive cutting, vision-based sorting, and battery removal to enable scalable automated recovery of high-value components. Mashhadi et al. \cite{mashhadi2017optimal} propose a data-driven sorting framework for remanufacturing that uses product life-cycle data to estimate a reusability index, cluster returned products by quality, and determine optimal recovery decisions such as refurbishment, remanufacturing, or recycling. Zhang et al. \cite{zhang2025predictive} propose a predictive repair management framework based on Transformer and online learning, using heterogeneous lifecycle and historical maintenance data to classify repair duration and support adaptive repair planning in evolving environments.

In summary, recent advances in robotics, control, and AI have greatly expanded the scope of intelligent automation within the remanufacturing pipeline, from inspection and sorting to disassembly and component reprocessing. These technologies improve efficiency and consistency while helping systems handle uncertainty, product variability, and complex tasks. As a result, intelligent automation provides a strong foundation for scalable, flexible, and more sustainable remanufacturing systems.

\section{Emerging Techniques for Remanufacturing}

The previous sections examined how intelligent automation is reshaping EoL product remanufacturing, highlighting applications of robotics, control, and AI for different stages. In this section, we shift our focus to several emerging techniques that may drive the future of this field. We emphasize their core concepts to provide both conceptual understanding and practical insights for practitioners and researchers.

\vspace{-0.1in}
\subsection{Large Foundation Models}

Recent years have witnessed significant progress in the development of foundation models, which are pretrained on large-scale and diverse datasets to learn generalizable representations. Unlike task-specific AI models, foundation models capture broad knowledge from vast data and provide a versatile basis for many applications. Once trained, these models can be adapted to different downstream tasks through techniques such as fine-tuning and prompt engineering. This paradigm significantly reduces the need to train separate models from scratch for each new task and enables knowledge transfer between different domains and problem settings. In this section, we review several categories of large foundation models, including LLM, VLM, VLA, and world model (WM), that are reshaping, or expected to support, intelligent automation for remanufacturing in the near future \cite{kim2026modular}.

\subsubsection{\textbf{Large Language Model (LLM)}}

LLMs, such as GPT \cite{brown2020language},  LLaMA \cite{touvron2023llama}, and PaLM \cite{chowdhery2023palm}, are foundation models trained to process and generate human natural language, typically by predicting the next token in a sequence. Through large-scale pretraining on massive text data from books, articles, code, and web content, LLMs acquire broad linguistic knowledge, semantic reasoning ability, and a degree of procedural understanding that can be transferred to downstream tasks. Mathematically, LLMs model the joint probability of a sequence of tokens $\mathbf{x} = (x_1, x_2, \ldots, x_T)$, where $x_t$ is the $t$-th token in the sequence and $T$ is the total length. Given a sequence $\mathbf{x}$, LLMs estimate the joint probability $P(\mathbf{x})$, which is factorized as: $$P(\mathbf{x}) = P(x_1, x_2, \ldots, x_T) = \prod_{t=1}^{T} P(x_t \mid x_1, x_2, \ldots, x_{t-1}),$$ and the model, typically using a Transformer-based architecture \cite{vaswani2017attention}, learns $P(x_t \mid x_1, x_2, \ldots, x_{t-1})$, which is the conditional probability distribution over the next token given all previous tokens, by minimizing the negative log-likelihood of the training data. During inference, the model generates text by autoregressively sampling tokens from the conditional probability distribution, continuing until a stop token is produced or a predefined maximum length is reached \cite{minaee2024large}.

There are several common ways to adapt a pretrained LLM to specific tasks. Prompt engineering \cite{chen2025unleashing} is the lightest form of adaptation, which does not change the model parameters. In this case, the user provides carefully designed prompts that describe the task, format, constraints, and examples. It works well when the pretrained model already possesses sufficient relevant knowledge and only requires clearer task specification. On the other hand, parameter-efficient fine-tuning \cite{han2024parameter} adapts a pretrained model to a specific domain that may not have been covered during pretraining. Instead of full fine-tuning of large models, which can be expensive in terms of memory, computation, and storage, methods such as LoRA \cite{hu2022lora} are widely used for LLM adaptation. The key idea is to preserve the pretrained backbone while learning a lightweight task-specific module using data from downstream tasks, making adaptation more practical. On top of LoRA-based adaptation, human preferences can also be incorporated through reinforcement learning from human feedback (RLHF) \cite{ouyang2022training}, in which the model is further tuned using feedback derived from human preference comparisons so that its outputs better align with human desired behaviors.

For intelligent remanufacturing, the value of LLMs lies less in direct physical actuation and more in their ability to serve as high-level reasoning, communication, and knowledge-integration modules. One of the most immediate applications of LLMs in remanufacturing is knowledge-intensive process support and decision assistance. Besides the LLM-based disassembly task planners that map scene descriptions and human instructions into candidate removal sequences \cite{xia2025leveraging, yu2025rescheduling, erdogan2025intent, tong2026gnn}, recent work has shown that LLMs can also be used to augment optimal remanufacturing process design by supporting the selection of remanufacturing schemes and process parameters in complex multi-objective settings \cite{zhang2025llm}. In summary, LLMs are promising not only as language generation tools but also as interfaces that connect human intent, process knowledge, and automated planning in remanufacturing scenarios.

\begin{figure}[t]
    \begin{center}
        \includegraphics[width=0.45\textwidth]{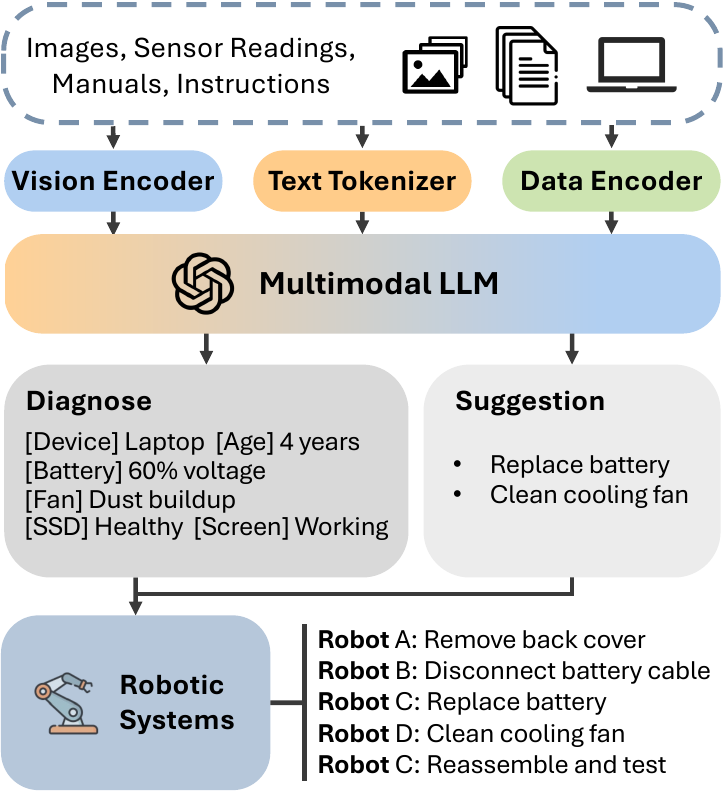}
        \caption{An Multimodal LLM-based laptop repair example, showing how LLM/VLM can contribute to EoL product remanufacturing. 
        }
    \label{figure_battery} 
    \end{center}
    \vspace{-0.3in}
\end{figure}

\subsubsection{\textbf{Vision-Language Model (VLM)}}

VLMs are a class of multimodal foundation models designed to jointly process visual and textual information \cite{ghosh2024exploring, zhang2024vision}. With the emergence of numerous studies in recent years, the architecture of VLMs has undergone multiple evolutions \cite{li2025benchmark}. Early works, such as CLIP \cite{radford2021learning} and BLIP \cite{li2022blip}, employ a vision encoder and a text encoder for pretraining via contrastive learning, which pulls paired images and texts closer while pushing unpaired examples farther apart in the embedding space. Later approaches, such as LLaVA \cite{liu2023visual} and Flamingo \cite{alayrac2022flamingo}, replace the text encoder with an LLM for text understanding and integrate visual embeddings through projection layers or cross-attention. Some recent models treat all inputs as tokens within a single transformer \cite{zhou2024transfusion}. Despite architectural differences, the primary goal of VLMs is to learn a shared, aligned representation space for both visual and textual data. By training on large-scale datasets, VLMs map image features and text embeddings into this joint space, enabling them to understand, reason about, and generate content over modalities.

With the additional visual modality, VLMs are generally well suited for inspection and sorting applications, where they can associate visual observations with semantic concepts, support open-vocabulary recognition, and improve defect understanding. For example, industrial anomaly detection work has shown that multimodal large models can support tasks such as anomaly recognition, localization, categorization, and interactive querying \cite{jiang2024mmad}. Similarly, VLM has also been tested for few-shot defect detection on multiple product categories \cite{ueno2025vision}. Beyond inspection, VLMs have also shown promise for sorting-related applications such as waste recognition, where zero-shot and few-shot vision-language classification can help categorize diverse materials in data-scarce settings \cite{malla2025enhancing}.

Specifically in remanufacturing scenarios, VLMs have also demonstrated significant potential for adaptive disassembly task planning under uncertainty. Ude et al. \cite{ude2025challenges} propose a framework for EoL product battery removal, in which the VLM serves as a high-level module for action selection and reasoning. Given the current visual scene, symbolic state, prior action history, and retrieved reference examples, the VLM selects the next valid disassembly action from a predefined skill library by evaluating the success of the previous action, predicting the next action, and verifying whether the conditions for executing the predicted action are satisfied. Additionally, Fig. \ref{figure_battery} presents an example laptop repair system, illustrating how a multimodal LLM can support EoL product diagnosis and repair recommendation. In summary, as multimodal models continue to improve, VLMs are expected to support more autonomous, knowledge-driven, and adaptable remanufacturing workflows.

\subsubsection{\textbf{Vision-Language-Action Model (VLA)}}

Building on the success of LLMs and VLMs, VLAs have emerged in recent years to tackle language-conditioned robotic tasks, marking a significant step toward general-purpose embodied intelligence \cite{ma2024survey, zhang2025pure}. Compared to traditional robotic systems, which often separate perception, planning, and control into distinct modules, VLA models aim to learn a direct mapping from visual observations and language instructions to robot actions. This makes them a representative approach for visuo-motor control \cite{zitkovich2023rt, kim2024openvla, kim2025fine, black2024pi_0, intelligence2025pi_, intelligence2025pi}, as we dicussed in previous section. Although various architectures have been explored in the existing literature \cite{ma2024survey}, the general idea is to combine a visual encoder, a language model, and an action prediction module within a unified policy, which can be written as: $$
\mathbf{a}_{t:t+H-1} = \pi_\theta(\mathbf{o}_{1:t},\, l,\, \mathbf{a}_{1:t-1}),
$$where $\mathbf{o}_{1:t}$ denotes the observation history, $l$ is the language instruction, $\mathbf{a}_{1:t-1}$ represents previous action history at time step $t$, and $\mathbf{a}_{t:t+H-1}$ denotes a chunk of $H$ future actions generated. By pretraining on large-scale robot manipulation dataset, such as the Open X-Embodiment dataset \cite{o2024open}, VLA models can acquire broad priors over object affordances, motion patterns, and instruction-conditioned behaviors for diverse tasks, embodiments, and environments. In practice, these models typically require further adaptation through task-specific fine-tuning, where the pretrained backbone is specialized using a smaller set of downstream robotic data collected in the target environment, either via full parameter tuning or through parameter-efficient methods such as LoRA.

For remanufacturing applications, many operations, such as prying, unscrewing, connector extraction, and insertion, are highly contact-rich and depend on subtle state transitions, including force and compliance, that may not be fully observable through vision alone. Therefore, purely vision-language-conditioned action generation may be insufficient for robust execution in such tasks. To support such applications, additional modalities have been incorporated into the VLA framework \cite{yu2025forcevla, huang2025tactile}. ForceVLA \cite{yu2025forcevla} treats force as the first-class modality, leveraging a force-aware mixture-of-experts approach to fuse tactile feedback with visual-language embeddings, thereby greatly enhancing performance in contact-rich  manipulation. Besides incorporating tactile sensing as an additional modality in the network, TACTILE-VLA \cite{huang2025tactile} also integrates tactile feedback into the chain-of-thought reasoning process to enable replanning for precise and adaptive control. Instead of relying on real force sensors, FD-VLA \cite{zhao2026fd} introduces a force-distillation mechanism that predicts a latent force token from visual observations and robot states, enabling force-aware reasoning and contact-rich manipulation without requiring direct force measurements during deployment. Overall, VLAs offer a promising path toward more intelligent remanufacturing robots. However, realizing this potential will require future work to push the boundaries of how multimodal sensing, safe control, and task-specific adaptation can be effectively integrated within the remanufacturing domain.

\subsubsection{\textbf{World Model (WM)}}
Unlike VLAs, which map multimodal observations directly to robot actions, World Models (WM) \cite{ding2025understanding, zhang2025step}, as a potential path toward achieving general embodied AI, learn internal representations that model environment dynamics and action consequences. This approach shifts the reasoning logic from “Given what I see and the instruction, what action should I take next?” to “If I take this action, what will happen next?” Besides language model-based methods \cite{chen2025egoagent, hong20233d, zhen20243d} and video generation models \cite{du2023video, wu2024ivideogpt}, which can also be interpreted through the lens of world modeling, latent dynamic modeling represents another promising technical path within WMs. Typically, high-dimensional observations are encoded into compact latent states using a variational autoencoder, and the temporal evolution of these latent representations under action is predicted using recurrent neural networks (RNNs) or Transformers \cite{lecun2022path, hafner2019dream, hafner2019learning}. Although WM-based methods remain largely unexplored in the remanufacturing and robotic disassembly domains, their strong performance in general control and manipulation tasks makes them a promising candidate for enabling intelligent remanufacturing automation \cite{li2026learning}.

\vspace{-0.1in}
\subsection{Human-Robot Collaboration (HRC)}

The significant uncertainty in remanufacturing, arising from variations in product geometry, wear, component quality, and unexpected defects, makes fully autonomous operation extremely challenging. Although robots excel at precision, repeatability, and physically demanding tasks, they lack human-level intelligence at the current stage. On the other hand, humans are highly effective at reasoning under uncertainty, recognizing subtle cues during inspection, and adapting to unforeseen situations. Human-robot collaboration (HRC) therefore offers a natural way to combine their complementary strengths: humans provide adaptive decision-making and contextual understanding, while robots perform repetitive, high-precision, or force-intensive operations. This synergy not only improves efficiency and accuracy but also enables remanufacturing systems to handle a wider range of products and scenarios than either humans or robots could manage alone. Depending on the interaction mode, HRC systems in remanufacturing can be broadly categorized into four levels \cite{jahanmahin2022human, dhanda2025reviewing}.

\subsubsection{\textbf{Coexistence}}

At the lowest level, coexistence refers to scenarios where humans and robots work in the same cell without physical barriers but on different tasks or objects, with no direct interaction \cite{magrini2020human}. Since they are not intentionally sharing the same task or object, safety demands are lower than in closer forms of HRC, mainly requiring safe workspace organization through monitoring, separation-distance control, and task scheduling. In remanufacturing, this mode suits settings where robots handle repetitive, structured operations such as pre-programmed disassembly, while humans perform separate inspection, diagnosis, or material-handling tasks in the same cell. Although coexistence offers limited collaboration, it is often among the easiest modes to deploy and can still improve system efficiency by enabling parallel human and robotic work within a shared production environment.

\subsubsection{\textbf{Sequential Collaboration}}

A second level is sequential collaboration, where the human and robot work on the same product or workflow in turn \cite{scimmi2021practical}. Unlike coexistence, their actions are linked through a shared workpiece or process, but unlike parallel collaboration, they operate at different times as ordered steps. Because the interaction is temporally separated, sequential collaboration generally requires less coordination than real-time teamwork. In remanufacturing, this mode is well suited to staged processes, for example when a robot carries out structured operations like unscrewing or cleaning before a human performs diagnosis or quality assessment. Therefore, it represents a practical intermediate level of HRC, offering tighter integration than coexistence while remaining easier to implement than fully synchronized collaboration.

\subsubsection{\textbf{Parallel Collaboration}}

A higher level is parallel collaboration, in which the human and robot work simultaneously on different but related subtasks of the same job to improve efficiency \cite{huang2021experimental, tian2023optimization}, such as performing unscrewing operations at the same time on the same EoL desktop. This requires tighter temporal coordination and greater workspace awareness. Unlike sequential collaboration, where the task is handed over from one to the other in turn, parallel collaboration involves concurrent activity in close proximity. This means that both agents need to be synchronized so that their actions interfere with each other as little as possible and the overall workflow proceeds smoothly. As a result, the main requirements extend beyond task allocation to include real-time coordination, mutual awareness of motion and intention, and more careful safety management within a shared workspace.

\subsubsection{\textbf{Full Collaboration}}

Finally, full collaboration describes tightly coordinated joint execution of the same task \cite{huang2020case, lee2022task, tian2025real}, similar to human-human teamwork, and may even involve direct physical interaction. At this level, the human and robot no longer simply share a workspace or divide subtasks, but actively contribute to the same operation at the same time toward a common immediate objective. Their actions are highly interdependent, meaning that the behavior of one agent directly influences the other throughout task execution. As a result, full collaboration requires the highest degree of coordination among all HRC levels, including real-time motion prediction and intention recognition. Although it is the most technically demanding mode to implement, it is particularly well suited for tasks requiring high flexibility, adaptability, and close human-robot coordination, which are exactly the qualities needed in remanufacturing.

These four levels reflect increasing degrees of interaction, coordination, and shared responsibility between human and robot, and together illustrate how HRC can be adapted to different remanufacturing requirements and operational complexities. In order to reach higher levels of collaboration in practice, disassembly automation systems must possess several key capabilities, including reliable perception of human workers, understanding and prediction of human intentions, adaptive task allocation, human-aware motion planning, and safe physical human-robot interaction. These topics have been extensively discussed in the previous section, where the underlying algorithms and system design considerations are reviewed in greater detail.

\vspace{-0.1in}
\subsection{Digital Twins and Simulation}

Remanufacturing differs from conventional manufacturing because discarded products exhibit significant uncertainty in quality, geometry, and wear condition. This variability makes process planning, inspection, disassembly, and repair considerably more complex than in traditional manufacturing systems. In this context, digital twins and simulation offer effective virtual, data-driven tools for modeling, predicting, and optimizing remanufacturing operations in both online and offline environments \cite{chen2021digital, wang2019digital}.

\subsubsection{\textbf{Digital Twins}}

A digital twin can be understood as a dynamic virtual representation of a physical asset, process, or system that continuously exchanges data with its real-world counterpart. It evolves over time by integrating real-time sensor data, historical records, operational parameters, and predictive analytics. This continuous synchronization allows the digital twin to reflect the current state, past behavior, and potential future conditions of the physical entity. As a result, digital twins enable more accurate monitoring, diagnosis, and decision-making across different stages of the product lifecycle \cite{jones2020characterising}. Typically, a digital twin consists of several core components, including a physical entity, an information model that abstracts the specifications of the physical object, a communication mechanism that enables bidirectional data exchange between the digital twin and its physical counterpart, and a data processing module that extracts and integrates heterogeneous multi-source data to construct a real-time representation of the physical system \cite{lu2020digital}.

In the remanufacturing domain, digital twins can support several key functions throughout the process. For example, they can support core tracking, where each returned product is assigned a digital identity and lifecycle record \cite{wang2019digital}. They can also assist in damage diagnosis and repair planning, particularly for high-value products \cite{ghorbani2022construction}. In addition, digital twins can facilitate decision-making on recovery strategies, such as choosing among reuse, remanufacture, recycling, or disposal \cite{kerin2023optimising}. Furthermore, they can support disassembly and logistics planning \cite{kerin2023generic}, especially reactive planning when real-time changes occur \cite{streibel2025integrating}.

\subsubsection{\textbf{Simulation}}

Simulation plays a complementary role by allowing manufacturers to evaluate strategies and algorithms in a virtual environment offline before deployment \cite{zhao2020sim, mourtzis2020simulation}. Its importance is especially pronounced in robotic remanufacturing, where tasks such as inspection, disassembly sequence planning, motion planning, and manipulation are characterized by high uncertainty and strong dependence on the condition of end-of-life products. By enabling these tasks to be tested and refined offline, simulation helps reduce the risk of collisions, control failures, and inefficient process execution.

In addition to offline validation, simulation provides an effective environment for training and benchmarking reinforcement learning methods. Control policies can be developed in simulation through trial and error before transfer to physical systems, thereby reducing the likelihood of costly or unsafe failures during real-world operation \cite{liu2020real, torne2024reconciling, wagenmaker2024overcoming}. Simulation also lowers the cost and complexity of data collection by generating large amounts of training data without repeated experiments on real hardware \cite{park2025dart, fishman2023motion}. Furthermore, it can support operator training by providing a safe and controlled environment for learning disassembly procedures and how to collaborate with robots during remanufacturing \cite{maddipatla2025vr}.

\section{Challenges and Opportunities}

Although substantial progress has enabled varying degrees of intelligent automation for different stages of remanufacturing, several barriers continue to hinder its large-scale deployment. These barriers arise from the fundamental characteristics of remanufacturing, including high product variability, incomplete or uncertain product information, unstructured operating environments, and the complexity of human-robot interaction.

A major challenge is \textbf{\textit{the limited availability and quality of data}}. Many intelligent, learning-based methods rely on large volumes of labeled or unlabeled data for training, yet acquiring such data in remanufacturing settings is often expensive, labor-intensive, and time-consuming. This difficulty is further amplified by the wide diversity of product models, usage histories, fault types, and damage conditions encountered in practice. Simulation and digital twin technologies can help alleviate this issue by generating synthetic data; however, discrepancies between simulated and real operating environments often limit the generalization and transferability of trained models, highlighting the need to address sim-to-real gaps effectively. Beyond synthetic data generation, further research on transfer learning and domain adaptation is also needed to address data scarcity in remanufacturing.

Another challenge in intelligent remanufacturing is \textbf{\textit{the generalization capability of learning-based methods}}, owing to the significant uncertainty and heterogeneity of EoL products. In practice, even products of the same model may differ substantially. As a result, the data distribution encountered during deployment often deviates from the training distribution used to develop perception, planning, or control algorithms. Such distribution shifts can significantly degrade model performance, making learned policies or predictive models less reliable when confronted with unseen products, novel damage patterns, or unstructured operating conditions. Therefore, research on out-of-distribution robustness, uncertainty awareness, few-shot and other data-efficient learning methods, as well as online and test-time adaptation, is essential for enabling reliable intelligent automation in future remanufacturing systems.

Another critical issue is \textbf{\textit{safety in the integration of humans and robots within shared remanufacturing workspaces}}. Collaborative robots offer the potential for flexible task allocation by combining human adaptability with robotic precision and repeatability. However, safe and efficient collaboration depends on reliable human motion prediction, continuous real-time perception, and advanced robot control strategies that can respond and adapt to dynamic changes. At present, robots still do not possess human-level intelligence and therefore cannot yet serve as fully reliable collaborators in complex and dynamic remanufacturing environments. Future work is needed on intelligent decision-making, shared autonomy, and adaptive interaction strategies to enable safer, more reliable, and more practical HRC in intelligent remanufacturing systems.

Regarding research work development, one important challenge is \textbf{\textit{the lack of standardized benchmarks and evaluation protocols}}. The current advancement of intelligent remanufacturing is hindered by the absence of widely accepted benchmark tasks, unified experimental settings, and consistent performance metrics. At this stage, fair comparison across different studies remains difficult, and reproducibility is often limited by case-specific setups and evaluation procedures. To address this issue, the research community should develop shared benchmark problems, representative test platforms, and common evaluation frameworks tailored to remanufacturing applications. Such efforts would support more rigorous assessment of intelligent methods, improve the comparability of research outcomes, and accelerate the transfer of new technologies from laboratory studies to industrial practice.

Besides the technical challenges, achieving \textbf{\textit{cost-effective deployment}} of intelligent remanufacturing systems is also a critical and highly practical issue. In the future real-world remanufacturing factories, intelligent automation typically relies on the integration of multiple sensors, robotic platforms, computing hardware, and software frameworks, which often leads to considerable system complexity and high implementation costs. Therefore, a key requirement for future large-scale deployment is to develop intelligent systems that not only deliver strong technical performance, but are also economical, easy to reconfigure for different production requirements, and simple to maintain.

Despite these limitations, intelligent automation still holds significant promise for improving sustainability and economic performance in circular manufacturing. Continued advances in the relevant research areas would enhance the reliability, adaptability, and practical deployability of intelligent remanufacturing systems, thereby supporting higher recovery efficiency, reduced material waste, and longer product lifecycles across a wider range of industrial scenarios.

\section{Conclusion}

Remanufacturing has emerged as an important strategy for enabling sustainable and circular manufacturing by extending product lifecycles and recovering valuable materials and components. However, the inherent uncertainty associated with EoL products, including variations in geometry, wear condition, and structural integrity, poses significant challenges for traditional automation approaches. These challenges have motivated increasing research interest in intelligent automation technologies that can enhance the flexibility, adaptability, and efficiency of remanufacturing systems. This paper has reviewed the existing literature on intelligent automation within various stages of remanufacturing, such as disassembly, inspection, sorting, and component reprocessing, and has further examined several emerging techniques with the potential to guide future research in this field, highlighting how robotics, control, and AI are jointly reshaping the remanufacturing process. Despite these advances, several challenges remain before intelligent automation can be widely deployed in remanufacturing systems. Issues such as high product variability, limited data availability and model generalization, and safe human-robot interaction continue to limit the scalability and robustness of current solutions. Continued interdisciplinary collaboration between academia and industry will be essential for translating these technological advances into practical remanufacturing applications.

\bibliographystyle{IEEEtran}
\bibliography{ref}{}

\end{document}